# Enhanced Evolutionary Symbolic Regression Via Genetic Programming for PV Power Forecasting

Mohamed Massaoudi[1,2], Ines Chihi[3], Lilia Sidhom[3], Mohamed Trabelsi[2,4], Shady S. Refaat[2], Fakhreddine S. Oueslati[1]
[1]Unité de Recherche de Physique des Semi-Conducteurs et Capteurs, Carthage University, Tunis, Tunisia
[2]Department of Electrical and Computer Engineering, Texas A&M University at Qatar, Doha, Qatar
[3]LAboratory of Research in Automation (LA.R.A), National Engineering School of Tunisia, Tunisia
[4]Department of electronics and communications engineering, Kuwait College of Science, Kuwait
{mohamed.massaoudi, shady.khalil}@qatar.tamu.edu;{Ines.Chihi, Lilia.Sidhom}@enib.rnu.tn;
m.trabelsi@kcst.edu.kw; fakhreddine.oueslati@fst.rnu.tn

*Abstract*—Solar power becomes one of the most promising renewable energy sources over the years leading up. Nevertheless, the weather is causing periodicity and volatility to photovoltaic (PV) energy production. Thus, Forecasting the PV power is crucial for maintaining sustainability and reliably to grid-connected systems. Anticipating the energy harnessed with prediction models is required to prevent the grid from any damage coming from every slight disturbance. In this direction, various architectures were suggested to predict the ambiguous behavior of meteorological data. Within this vein. Genetic algorithm (GA) presents a robust solution for nonlinear problems. The success of GA presents a source of motivation to scientists and engineers to develop a variety of sub-models that imitate the same Darwinian type-survival of the fittest strategy approach from GA propriety. However, during the training process, the later face an issue with missing the optimal solutions due to the existence of a local minimum.

Following that regard, this paper provides an accurate PV power forecasting one month of PV power using a hybrid model combining symbolic regressor via Genetic programming and artificial neural network. The features inputs used in the process are only the solar irradiation and the historical solar power data. The application of the said model on an Australian PV plant of 200 kW offers a low mean absolute error equal to 3.30 and outperforms the state of art models.

## I. INTRODUCTION

In the few recent years, the world has witnessed exponential attention towards alternative energy sources. According to National Renewable Energy Laboratory (NREL), the PV stations have recorded a total of 509 GW-DC during the last months of 2018 with a rise of 102 GW-Dc from the previous year[1]. This motivation for making this transition policy comes as a consequence of the increasing rate of air pollution and the lack of traditional sources in the incoming next years[2]. Moreover, abundant solar energy has many applications such as thermal and electric generation[3][4]. Photovoltaic panels are widely used to harness the maximum energy from the sun to transform it into electricity. However, PV generators are sensitive to weather conditions[5]–[7]. Certainly, climate parameters are continuously changing during the day. While at night, no PV energy is produced due to the lack of solar irradiance. From that standpoint, forecasting comes to determine the next PV power during many time steps ahead[7]–[9]. Precise forecasts are vital for preventing PV plants from a serious problem occurring in sudden damages[10]. This issue diminishes the penetration of grid-connected PV systems into a public utility.

Forecasting models provide a safe unit commitment and fast protective dispatches to the grid utility[11]. The latter are classed into three categories pending on the forecasting horizon. They are short, medium- and long-term prediction[12], [13]. Time series prediction is done through an extensive analysis of the weather patterns such as the temperature and the irradiation[14]. The forecasting process is typically done through numerical weather prediction (NWP)[15]. Generally, Markov models are used due to the fact that the actual predicted power is not affected by the previous prediction[16]. A variety of models are introduced using domain knowledge for the estimation of the future generated power[17]–[21]. On the other side, physical models are able to estimate the current PV power but with a lower precision[22], [23]. The stochastical behavior of the metrological data is predicted using statistical models such as Autoregressive–moving-average (ARMA) and Autoregressive–moving-average with exogenous inputs (ARMAX) to determine indirectly the PV power[24]. Moreover, Vagropoulos et al. evaluate Seasonal Autoregressive integrated moving average (SARIMA), SARIMA with exogenous inputs (SARIMAX) and Modified SARIMA for Short-Term PV Generation Forecasting[24].

For the performance investigation of artificial intelligent hybrid models on PV power prediction, Ayoub Fentis et al. investigated a bench of non-linear auto-regressive models namely feed-forward (FFNN) and least squares support vector regression (LSSVR) and compared them to the non-linear autoregressive models with exogenous inputs NARX[24]. In addition,

Genetic algorithm (GA) as one of the efficient methods proves its capabilities in forecasting through different applications. Firstly introduced by Holland in the 1970s[25], [26], GA imitates the biological evolution by multiple replications of its units[27]. The process mechanism is made by individual selection, mutations, and crossover[27]. This architecture provides the primary insights for developing many powerful models such as machine learning (ML), deep learning (DL), and intelligent search[27]. ML models as a part of GA provide an accurate result[27]. Muhammad Naveed Akhter et al. profoundly reviewed Artificial Neural Network (ANN), Support Vector Machine (SVM) and Extreme Learning Machine (ELM) techniques for smart grid systems[24]. At this stage, the insight of increasing the accuracy while using deep ML models with multiple layers is the key motivation for deploying these architectures in PV power prediction. Kejun Wang et al. compare deep learning models namely Long Short Term Memory(LSTM) and Convolutional Neural Network(CNN) and a hybrid model combining the aforesaid algorithms in photovoltaic power prediction[24]. The criteria for choosing the suitable model depends on its complexity, forecasted horizon and accuracy rate. Ensemble models and hybrid architectures are frequently used to ameliorate one or more aforementioned proprieties. From that standpoint, Yuxin Wen et al. proposed a hybrid model from Wavelet Transform (WT), Radial Basis Function Neural Network (RBFNN) and Particle Swarm Optimization (PSO) to enhance the effectiveness of the predictor [28]. While Yordanos Semero et al. combine GA, PSO and Adaptive Neuro-Fuzzy Inference Systems (ANFIS) with the goal of reducing the error value[29]. Following that regard, the contributions of this paper are resumed in three folds:

1. Firstly, the is a survey on Symbolic Regression model, a Multilayer Perceptron(MLP) algorithm, as well as genetic programming.
2. Then, feature engineering and domain knowledge design are deploying through a serious feature analysis.
3. The next part will introduce the proposed hybrid algorithm from SR and MLP with a case study on the PV energy.
4. The proposed model is analyzed and compared to individual models. A fair assessment is provided via numeric scores performance and graphic results that illustrate the behavior of the hybrid model.

## II. SYMBOLIC REGRESSION

Contrarily to the majority of machine learning models that propose a predefined function with prior assumptions for the fitness process, Symbolic regression (SR) build the mathematical expression that suitably fits the proposed database during the training stage[24]. The fitness of the symbolic regression is pending on simplicity and accuracy. SR suggests genetic programming (GP). i.e. Thus, the latter is considered as an evolutionary algorithm. The representation of SR is in the shape of trees. Fig. 1 presents a general idea of how the symbolic function is built through SR.

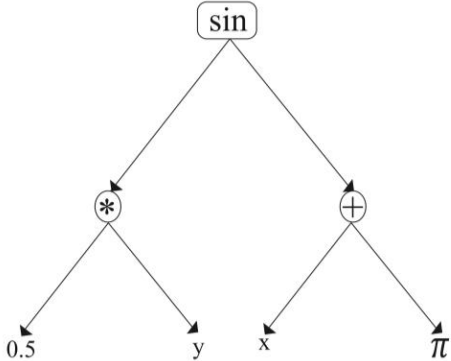

Fig. 1. Binary Genetic Tree Programing representation of the function $f(x,y) = sin(x + \pi + 0.5y)$

The internal nodes present the beginning of a mathematical operation ended by a leaf. The methodology is investigating the dataset parameter patterns with a variety of mathematical operations with analytic functions and state variables in the training stage. From a hierarchical form, the inputs are fed to the system and some fitness functions are constructed in the first iteration. By a random alteration named mutation and swapping parts (crossover), the application of the latter generates an error value. Then, gene duplication is done to produce the descendants' offsprings. The latter replaces the first generation to give birth to a final symbolic function. The aim is generating a novel individuals only from stronger genes respecting the Darwinian type-survival of the fittest strategy[30]. Mutations are done randomly with the aim of reducing the rooted mean square value (RMSE). When the error is reaching the minimum threshold, the symbolic function is fixed and the training part is finished to pass to the evaluation process. Obviously, only supervised problems are well-performing with SR since the database is the primary responsibility for shaping the symbolic function of the algorithm. The strength of the aforementioned model is coming from its propriety to let the dataset itself choose the best function that matches the lower RMSE. The crucial parameters of SR the population size, generation, stopping criteria and the mutation point.

Although the cited features of SR, the main disadvantage of the latter is the large search space with an infinite generation that presents an accurate result. The searching process is time-consuming with a variety of local minimums. Thus, the model risk of being tricked with a faulty suboptimal solution.

## III. DEEP MULTI-LAYER PERCEPTRON

Multi-layer perceptron is a deep feedforward artificial neural network containing essentially an input, hidden and output layers. In MLP, the information has the propriety of a unidirectional propagation. According to Minsky and Papert[29], every perceptron is activated through a non-linear activation function e.g. Sigmoid, Rectified Linear Unit (ReLU), hyperbolic tangent function (tanh) and Normalized exponential function (Softmax)[31]. The equations of the nonlinear activation functions are given in Eq.1-4.

$$\text{Sigmoid}(x) = \frac{1}{1+\exp(-x)} \quad (1)$$

$$\tanh(x) = 2\sigma(2x) - 1 \quad (2)$$

$$\text{ReLu}(x) = \max\{0, x\} = \begin{cases} x \text{ if } x \geq 0 \\ 0 \text{ if } x < 0 \end{cases} \quad (3)$$

$$\text{Softmax}(x)_j = \frac{e^{x_j}}{\sum_{k=1}^{K} e^{x_k}} \quad (4)$$

Each connection has a specific weight that indicates its importance. While every neuron has a characteristic value named bias. During the propagation process, each input $x_i$ is multiplied by its connection weight $w$ and summed with the bias value $b$ presented in Eq. 5.

$$y = w*x + b \quad (5)$$

The nonlinear activation function is applied to the residual, Then, the latter value is spread to the next layer. The same process is repeated until having the final yield from the output layer. Eq. 7 will explain more the mathematical function.

$$y = \varphi(\sum_{i=1}^{n} w_i x_i + b) = \varphi(w^T x + b) \quad (6)$$

With $y_{output}$ the output of the system and $\varphi(.): R \rightarrow R$ is the nonlinear activation function. The structure of the MLP is illustrated in Fig. 2 where the nodes are connected to each other via weighted linkers.

The last step in the training process is backpropagation. At this stage, the weights and the bias are tuned according to the loss function. This parameter presents the difference between the actual and the predicted values. The gradient-based optimization algorithm in every iteration identifies the learning rate of the system and leads to minimize the error to converge to a lower value. Since the number of hidden layers and neurons is high, MLP presents the primary form of deep learning. This model has addressed many supervised problems from different applications involving Natural language processing (NLP), regression and classification programs. The existence of multiple layers for a large number of neurons is making the architecture more efficient and outperform the benchmarked algorithm despite its simplicity. The use of the MLP for one-day prediction of PV power via forecasting the solar irradiance gives an accuracy of 99% [24]. This model is able to handle nonlinear problems with high effectiveness. Nevertheless, the latter suffers from redundancy in high dimensions and the sensitivity to the inputs scaling. Hyperparameters tuning including the number of neurons and layers, the activation function type and the initial bias and weights is an essential step for straightening the model architecture and guarantying a faster convergence to the desired target.

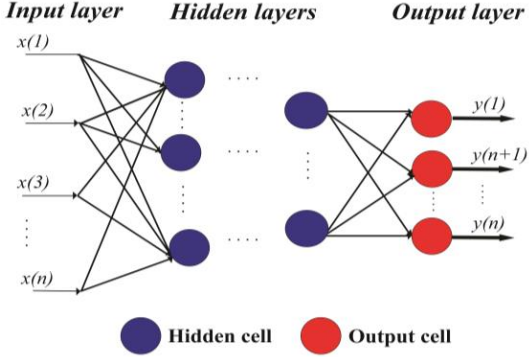

Fig. 2. Feed forward neural network architecture

IV. GENETIC PROGRAMMING

Genetic programming (GP) and genetic algorithm are very similar in a manner that identifying the difference between them is quite tricky. Genetic programming firstly introduced by J. R. Koza[32] presents its architecture in a set of tree structures. The nodes are the operation functions that constricts the final function. While the form of GA is different in a linear structure with a number of sub-branches. From that standpoint, genetic programming is having better flexibility with less invalid states. The hierarchical form prevents the used operators from the precedence. GP is a sort of supervised computer guidance to find solutions for high-level complicated problems with the vital need for machine intelligence[33]. Expert systems (ES) defined by Feigenbaum as a hyper-intelligent computers require such a strong architecture in decision making[34]. The expertise of human reasoning is cloned via GP through heuristic rules to solve narrow domains. In this paper, SR is simply a GA developed by evolutionary algorithms. GP is used in nonlinear problems that require a better domain understanding and extensive machine intelligence such as electronic circuits. These problems involve interpretation, prediction, diagnosis; planning, monitoring, debugging and control. GA works with the aid of a bench of commands namely automatically defined recursions (ADR), automatically defined loops (ADL), automatically defined functions (ADF) and automatically defined iterations (ADI)[33].

V. PROBLEM FORMULATION

PV project planners and investors face a serious issue when studying the feasibility of their projects. Since the weather is intrinsically volatile. The solar power generation seems to be hard to expect. The fears from the intermittent nature diminish the expansion for a wilder rate. Researchers try to cover this weakness through different techniques. Solar trackers follow the sun during the day. Thus, the energy is relatively maintained at a level. The non-linearity of the PV power parameters is a participant in this behavior. This is due to the fact that the PV power is taking into account meteorological data which are not necessarily proportional to each other. Eq .8 explains more the relationship between the weather parameters.

$$P_{PV} = V_{pv} N_p \left( \frac{(I_{sc} + K_i(T - T_{ref}))G}{G_{ref}} - I_d - I_{sh} \right) \quad (7)$$

With $P_{pv}$ is the power, $T$ is the cell temperature, $G$ is the irradiation, $T_{ref}$, $G_{ref}$, $I_d$ $K_i$, $I_{sc}$, $V_{pv}$ and $N_p$ are the referent temperature, referent irradiance, diode current, short circuit current/temperature coefficient, the shunt current, PV voltage and number of parallel cells respectively. The PV power is a summation with nonlinear parameters In a continuous variation during the day as shown in Fig.3. for a single module with the Australian weather. Thus, the power output should be anticipated to make a clear strategy on the manner of extracting the maximum benefits of the infinite solar energy. This paper provides an efficient method with one month as a solution for that matter.

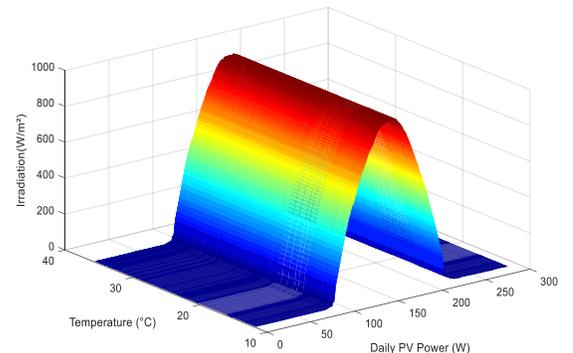

Fig. 3. PV power generated in one day in function of the irradiation and the temperature

## VI. PROPOSED MODEL

The proposed model is a combination of GA via SR and a deep Feedforward model. The aim is building an ensemble of subtrees with heterogeneous units. The mutation is done through finding the local minimum between these two models which take into consideration the residual of their offsprings. During the simulation of the PV power with GA undividially, it has been noticed that the majority of the predicted points are above the real values. The key insight for developing these blending models comes from decreasing the error with a model in which the predicted values are underestimated. The dataset had a severe feature selection. So, the predictor is relying on lesser features in medium-term prediction for one month. Then, the solar irradiation and the previous PV power are fed simultaneously to the SR and the MLP. The output of the two systems is averaged to generate the final result. The proprieties of the tow algorithms are merged to build a strong predictor. Unlike ensemble models that combine homogenous models, the suggested estimator combines two heterogeneous tree structures in which the first uses mathematical operators and the second uses multiple neurons. The last layer uses a single operator to get the final output. By this method, transfer learning is preventing the system from noise and losses.

The dataset passes to a feature engineering process to eliminate the missing and faulty values coming either from a sensor damage or record errors. Then, Deep Multilayer Perceptron and Symbolic Regressor are trained individually. The output of these tow predictors will be averaged via the voting technique. Fig. 4 present a detailed description for building the predictor. Note that the proposed predictor is assumed to reach the nearest forecast to the ground truth. Support vector machine (SVM) and K-nearest neighbors (KNN) used the same mechanism during the prediction. We adopt that the proposed method can protect the symbolic regressor from falling into a local minimum. With extending the trees to involve MLP and symbolic functions, the prediction reliability will be enhanced. since the predictors do not have the same proprieties, the fusion can make the latter complementary. By reassembling them together, the final predictor has better robustness. Taking as example Random forest proposed by Kam [35], the combination between an ensemble of Trees contribute a supplement strength to the system. The next algorithm describes more the proposed architecture.

**Algorithm1 : Hybrid model**

❖ **Input:**
1. Data acquisition $X_i = \{X_i, .., X_n\} = \{IR; T; RH; WS; PV_{power}\}$
2. Feature selection $X_i = \{IR_i, PV_{power}\}$

❖ **Output:**
1. Data splitting to 80% for training $\{X_{train}, Y_{train}\}$, 20% for testing $\{x_{test}, y_{test}\}$.
2. Hyper parameter optimization
3. Train Symbolic regressor (Model 1)
4. Use $x_{test}$ to predict $y_{predicted1}$ with Model 1.
5. Train MLP regressor (Model 2)
6. Use $x_{test}$ to predict $y_{predicted2}$ with Model 2.
7. Hybrid model mixture
   $y_{output} = avreage(y_{predicted1}; y_{predicted2})$
8. Assess the model with Cross Validation, scores metrics and simulation graphs

## VII. CASE STUDY

### A. Feature engineering

The aim of writing this paper resume in having a precise forecasts of the PV energy during one month. To achieve this goal, a complete database containing historical records of weather parameters is required. The Australian KASC, Alice Springs site has complete data for two successive years of 2018-2019[36]. The latter contains the temperature, the relative humidity(%), the horizontal and diffuse irradiation(W/m²), the wind proprieties in terms of speed(m/s) and direction (°) as well as the measured PV power(kW). This dataset includes nearly all parameters that can affect the PV plants with a time step equal to 5 minutes. The training set is fixed from the first of January 2017 till 31 December 2018. The testing will be focused on the month of January. Features collection shrink the database size to speed the training process and enhance the accuracy rate. Tow efficient feature selection methods are applied, namely Elastic Net and Extreme Boosting. These techniques are frequently used in attribute selection. The combination of these techniques helps in getting more reliable results since the mechanism of measuring the parameter magnitude is different. Fig.5 illustrates the variable selection results.

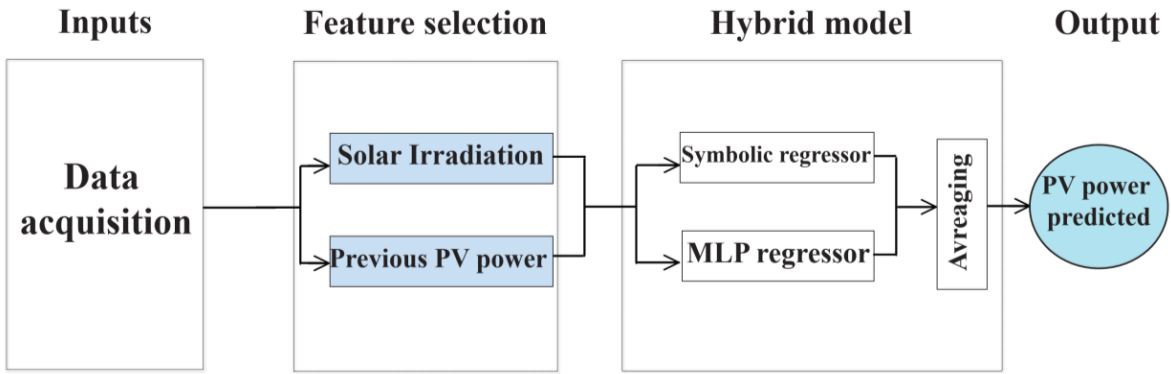

Fig. 4. Proposed model schema

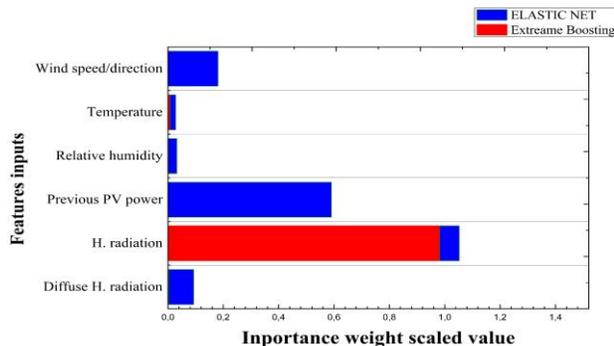

Fig. 5. Nonlinear correlation coefficient of attributes with the PV power

From Fig. 5, it has been noticed that the horizontal irradiation and the previous PV power have significant importance more than any other input. Extreme booting shows that the radiation as a crucial indicator of the current PV power. While Elastic Net method had given the previous PV power more importance. Note that the previous photovoltaic power presents the historical value from the same date, the same minute for the previous year. Fig. 6 and 7 illustrate the behavior of the aforementioned parameters for two entire years.

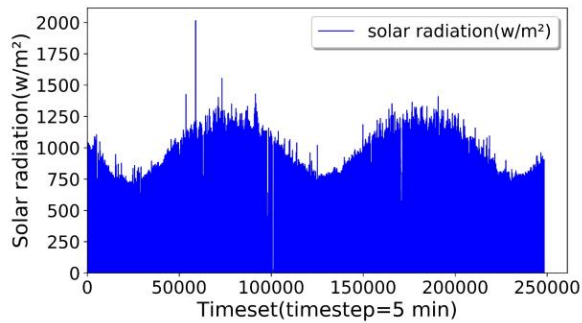

Fig. 6. Irradiation variance(w/m²)

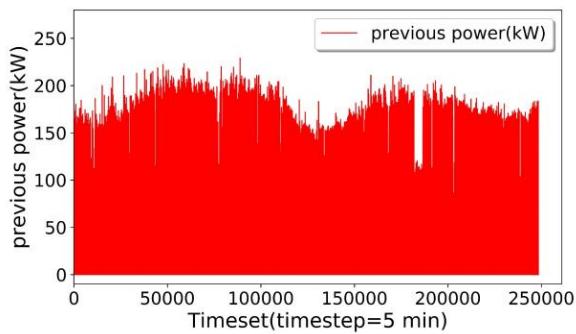

Fig. 7. Previous PV power(kw)

From features selection, a serious study of the causes of these results is made. This investigation is vital since Eq. 8 presents the temperature as a direct factor that enhances or reduces the PV model performances. While analyzing the relationship of the pattern between the PV power in one side and the chosen parameters from the attribute methods on the other side, It has been released that these input parameters are having the same shape as the final output. This appears clearly in figure 8 and Fig 9. Thus, in this study, the selected parameters are assumed sufficient for PV power forecasting. Note that all the predictors in this study will use this database for prediction.

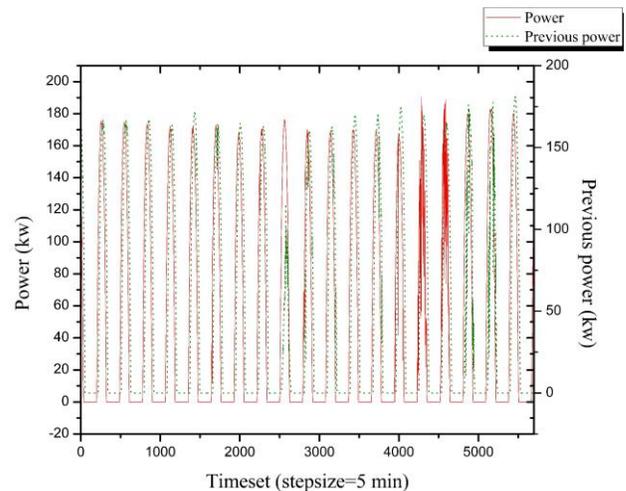

Fig. 8. Actual and previous PV power during correlation during January 2018-2019

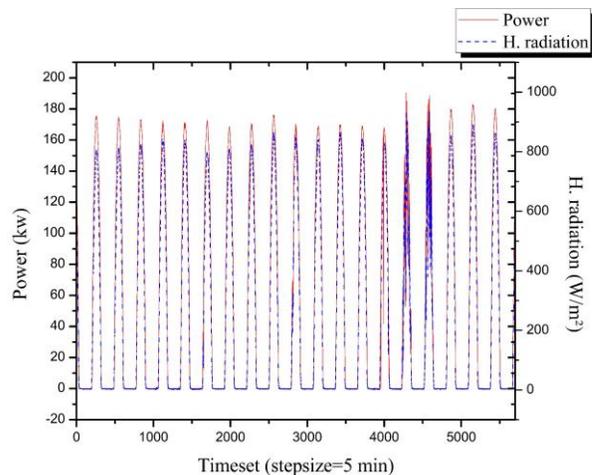

Fig. 9. PV power and horizontal radiation correlation during January 2018-2019

*B. Training and simulation results*

Symbolic regressor and MLP, as well as the mixture between them, are interpreted in this section. The data is pre-processed with the elimination of the missing and anomalous data. This step is crucial for better learning. Then, the resulted data is rescaled between 0 and 1. This unification gives a better understanding of the feature's behavior especially for MLP model. The modeling and the training part are done through Python programming language. The hyperparameters are selected for each model through a Randomized Search method. Thus, MLP has 500 hidden layers with 3000 iterations, while SR has 3000 population size and 15 generations and The evaluation process is done through 3 categories: simulation graphs, Cross-Validation, and score metrics. The selected error metrics include rooted mean square error (RMSE), mean absolute error (MAE) as well as the coefficient of determination (R). Eq. (8)-(10) presents the mathematical equations of these score parameters.

$$R^2 = 1 - \frac{\sum_{i=1}^{n}(\hat{y}_i - y_i)^2}{\sum_{i=1}^{n}(\overline{y}_i - y_i)^2} \quad (8)$$

$$RMSE = \sqrt{\frac{1}{n}\left(\sum_{i=1}^{n}(\hat{y}_i - y_i)^2\right)} \quad (9)$$

$$MAE = \frac{1}{n}\sum_{i=1}^{n}(\hat{y}_i - y_i) \quad (10)$$

The experimental results are taken from real-time series data. January 2019 is the testing month. The simulation is presented for one month and one day in Fig. 10 and 11 respectively.

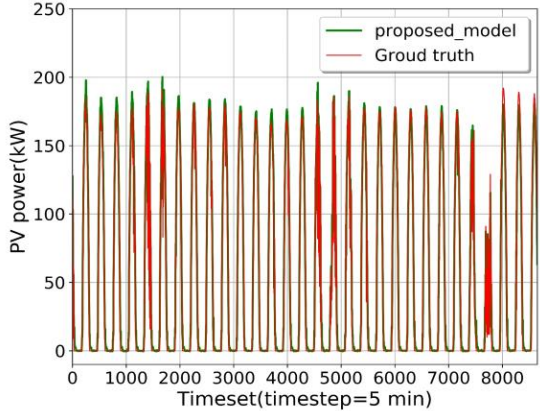

Fig .10. Previous PV power(kw)

Regarding Fig 10, the predicted value from the proposed model is highly accurate with a slight error value. The green shape presenting the forecasted PV power is identical to the real PV power in the majority of points. Moreover, the training time took 11 minutes which is considerably low. The suggested algorithm is well-performing with time series data and for a period of time that reaches one month. Cross-validation is applied to the to proposed model in Fig.12.

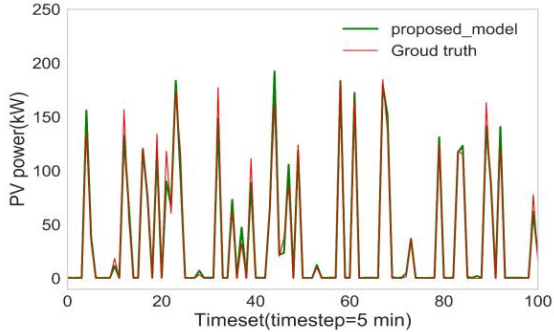

Fig. 11. Cross-Validation of the proposed method illustration

Cross-validation is an efficient method for ML model evaluation. It consists of splitting the rows in the training process from a k-folds of subdatasets. From figure 12 The proposed model with the green color is cross validated and compared to the real values. The maximum error measured is 30 kW. This presents a 13% of the PV power. Which is considerably high. While in the majority of the forecasts is low. This rate is varying over the timesteps. For investigating the enhancement rate of the proposed model from SR and MLP individually, Fig 12 is plotted. Furthermore, Fig 13 presents a zoomed shape to the model behavior compared to the single models.

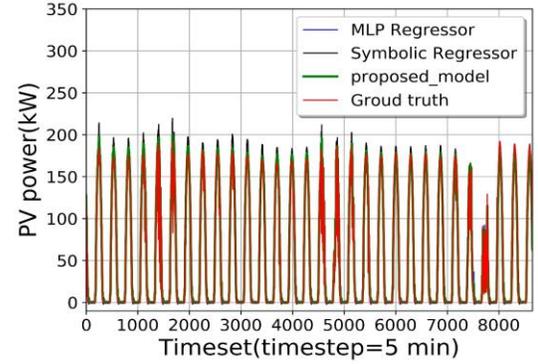

Fig. 12. Model comparison for PV power forecasting

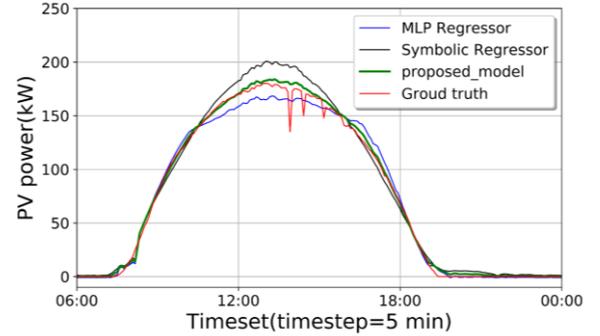

Fig. 13. PV power forecasted in one day (kw)

From Fig 12 and 13, it has been considerably remarked that the proposed method is outperforming SR and artificial neural network predictors separately. The transfer learning via the voted method has ameliorated the prediction accuracy in terms of the corresponding points between the ground truth and the real values. The hybrid tree combines tow sub-branches with an averaging interconnected leaf point. By using only, the irradiation and the previous PV power, the model generate a precise estimation. The suggested predictor prevents the system from overfitting and maintains a great efficiency during all the forecasted horizon. Table 1 presents the numerical score values of each predictor. Moreover, Fig. 14 and 15 show a comparative error result.

TABLE. 1. Score errors comparison

| Error | Symbolic regressor | MLP | Hybrid method |
|---|---|---|---|
| RMSE | 7.21 | 6.48 | 5.58 |
| MAE | 4.92 | 3.81 | 3.30 |
| R² | 98.85% | 99.07% | 99.31% |

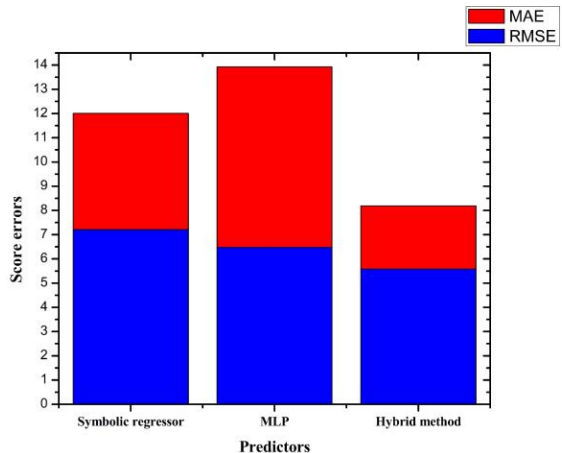

Fig. 13. MAE and RMSE Error comparison

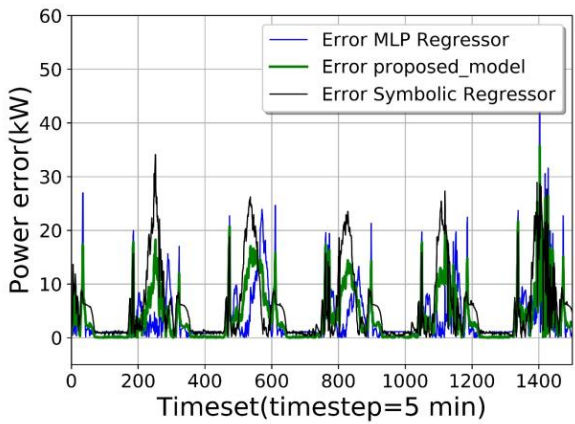

Fig. 14. Error difference

*C. Interperation*

The proposed model was verified through Fig 13 and 14. Table 1 records an accuracy equal to 5.58 in terms of RMSE and 3.30 in terms of MAE. The coefficient of determination is 99.31. These values lead that the enhanced SR is outperforming the separated models. The training data and the testing data are fed to the systems equally. Note that the hyperparameters are conserved for all the training process and the testing part. It should be mentioned that the system has a better efficiency more than the models separated. It should be mentioned that the final output is coming for averaging the individual predictors' results. More research on the variation of the contribution rate of each predictor for the aim of getting an optimum result is required and needs more investigation. Nevertheless, the proposed predictor still an efficient tool for maintaining the grid-connected PV system safe from any sudden disturbance.

## VIII. CONCLUSIONS AND OUTLOOK

This study suggests a hybrid method for forecasting the photovoltaic power according to Alice springs DCSK PV plants in Australia. The proposed method is a fusion of Symbolic Regression and a Multi-Layer Perceptron. The forecast horizon is one month using only two features specifically the horizontal irradiation and the historical annually photovoltaic power (for the same day and the same minute) for two years. The transfer learning via voting approach leads to an accurate forecast over one month. The RMSE is 5.58 while the MAE is equal to 3.30. Thus, it outperforms the benchmarked separate models with an enhancement of 22%RMSE and 13%MAE according to SR and MLP respectively. For each forecasted timestep, the proposed mixture creates an optimum by averaging the model's outputs.

Moreover, the efficiency of the proposed model was demonstrated through Cross-Validation and simulation figures. The graphs show a perfect match between the two trees architectures. This is shown clearly in Fig. 15 where the hybrid model estimation is generated with less error value. The advantage of using the aforementioned ensemble approach is its simplicity in implementation with The rapidity during the training stage. The suggested multimodal allows the MLP of using a lesser number of layers and neurons with maintaining the accuracy at its maximum. On the other side, in SR, the number of iterations is reduced. The said parameters are normally a time-consuming. Parallel computing has considerably decreased the computational cost to take only 11 minutes for two years of historical database with a timestep of 5 minutes in a LENOVO Ideapad 720S-15IKB i7 with 8 Cpu with Python 3.7 version. The elimination of minor important inputs from a feature selection process using Elastic Net and Extreme boosting has improved scientifically the training speed of the blending algorithm. Therefore, the latter is high performing in online forecasting with real-time implementation.

To sum up, the proposed hybrid model is highly recommended in PV power forecasting for one month since it has demonstrated its effectiveness through different tests. The reliability of this approach is counting on tow predictors which remarkably enhances the accuracy. The robustness of the blending model prevents the SR form converging to the local minimum with a complementary aid form the artificial neural network branch. Although the dataset was shrunk to include just tow features, the model is still suitable for medium-term forecasting while if the additional data was taken also in consideration, the accuracy will mostly have a higher value.

The proposed method will contribute to the grid utility in terms of unit commitment and economic dispatch. In addition, the application of the aforementioned multimodal on a real PV plant is still required to validate the proprieties on online training. Voting technique via machine learning models is the key success of the proposed approach. However, if the SR doesn't inversely follow the MLP, the error will dramatically increase. For that reason, the development of an indicator which indicates the sign and guides the mixture to an accurate forecast is needed to prevent the hybrid predictor from any mislead.

The future work of these study includes a serious investigation on the proportional contribution of each model to the domain knowledge with an online tuning targeting the minimum error with extending the prediction horizon up to several months.

## ACKNOWLEDGMENTS

The authors highly acknowledge the National Priorities Research Program (NPRP) from the Qatar National Research Fund (a member of Qatar Foundation) for the financial support of the. Also special thanks to Pr. Haithem Abu-Rub from Smart Grid Center Laboratory SGC who made this study possible. The findings achieved herein are solely the responsibility of the author(s).